\documentclass[conference]{IEEEtran}
\usepackage{color}

\usepackage{times}
\usepackage[latin1]{inputenc}
\usepackage{amsfonts}
\usepackage{balance}
\usepackage{amsbsy}
\usepackage{amssymb}
\usepackage[T1]{fontenc}
\usepackage{bm}
\usepackage{soul}
\usepackage{array}
\usepackage[linesnumbered,lined, algoruled]{algorithm2e}
\usepackage{tikz}
\usepackage{mfirstuc}
\DeclareMathAlphabet{\mathpzc}{OT1}{pzc}{m}{it}

\usepackage[acronym,shortcuts]{glossaries}

\usepackage{tikz}
\usetikzlibrary{arrows,%
                petri,%
                topaths}%
\usetikzlibrary{arrows}

\makeglossaries

\newacronym{ACK}{ACK}{acknowledge}
\newacronym{ASR}{ASR}{achievable sum-rate}
\newacronym{ARQ}{ARQ}{automatic repeat request}
\newacronym{AWGN}{AWGN}{additive white Gaussian noise}
\newacronym{BS}{BS}{base station}
\newacronym{BCC}{BCC}{broadcast channel with confidential messages}
\newacronym{BBU}{BBU}{base band unit}
\newacronym{CDF}{CDF}{cumulative distribution function}
\newacronym{C-RAN}{C-RAN}{cloud radio access network}
\newacronym{CSI}{CSI}{channel state information}
\newacronym{CSI-FB}{CSI-FB}{\ac{CSI} feedback}
\newacronym{CSIT}{CSIT}{channel state information at the transmitter}
\newacronym{DEC}{DEC}{decoder}
\newacronym{DFT}{DFT}{discrete Fourier transform}
\newacronym{ENC}{ENC}{encoder}
\newacronym{GC}{GC}{graph coloring}
\newacronym{GMP}{GMP}{Gaussian message-passing}
\newacronym{PGC}{PGC}{partial \ac{GC}}
\newacronym{RGMP}{RGMP}{randomized GMP}
\newacronym{VR}{VR}{variable-rate}
\newacronym{FR}{FR}{fix-rate}
\newacronym{HARQ}{HARQ}{hybrid automatic repeat request}
\newacronym{iid}{i.i.d.}{independent and identically distributed}
\newacronym{IR-HARQ}{IR-HARQ}{incremental redundancy HARQ}
\newacronym{LDPC}{LDPC}{low density parity check}
\newacronym{MIMO}{MIMO}{multiple input multiple output}
\newacronym{MI}{MI}{mutual information}
\newacronym{MRC}{MRC}{maximal ratio combining}
\newacronym{MMSE}{MMSE}{minimum mean square error}
\newacronym{MP}{MP}{message passing}
\newacronym{MT}{MT}{mobile terminal}
\newacronym{NACK}{NACK}{not acknowledge}
\newacronym{PDF}{PDF}{probability density function}
\newacronym{PLS}{PLS}{physical layer security}
\newacronym{PMF}{PMF}{probability mass function}
\newacronym{RRH}{RRH}{radio remote head}
\newacronym{SNR}{SNR}{signal to noise ratio}
\newacronym{SINR}{SINR}{signal to interference plus noise ratio}

\IEEEoverridecommandlockouts
\begin{document}

\title{Beamforming and Scheduling for mmWave Downlink Sparse Virtual Channels With Non-Orthogonal and Orthogonal Multiple Access}

\author{
	\IEEEauthorblockN{Alessandro Brighente and Stefano Tomasin} \thanks{This work has been supported in part by Huawei Technologies, Milan, Italy.}
	\IEEEauthorblockA{Department of Information Engineering, University of Padova  \\
                             via Gradenigo 6/B, 35131 Padova, Italy. \\ Email: brighent@dei.unipd.it, tomasin@dei.unipd.it }
                             }

\maketitle

\begin{abstract}
We consider the problem of scheduling and power allocation for the downlink of a 5G cellular system operating in the millimeter wave (mmWave) band and serving two sets of users: \ac{FR} users typically seen in device-to-device (D2D) communications, and \ac{VR} users, or high data rate services. The scheduling objective is the weighted sum-rate of both \ac{FR} and \ac{VR} users, and the constraints ensure that active \ac{FR} users get the required rate. The weights of the objective function provide a trade-off between the number of served \ac{FR} users and the resources allocated to \ac{VR} users. For mmWave channels the virtual channel matrix obtained by applying fixed discrete-Fourier transform (DFT) beamformers at both the transmitter and the receiver is sparse. This results into a sparsity of the resulting multiple access channel, which is exploited to simplify scheduling, first establishing an interference graph among users and then grouping users according to their orthogonality. The original scheduling problem is solved using a graph-coloring algorithm on the interference graph in order to select sub-sets of orthogonal \ac{VR} users. Two options are considered for \ac{FR} users: either they are chosen orthogonal to \ac{VR} users or non-orthogonal. A waterfilling algorithm is then used to allocate power to the \ac{FR} users.
\end{abstract}

\begin{IEEEkeywords}
5G; Beamforming; mmWave; Non-orthogonal Multiple Access; Scheduling.
\end{IEEEkeywords}

\glsresetall

\section{Introduction} 
 MmWave transmission systems operating with multiple users will pose many design challenges, related to the need of precise beamforming to overcome the strong attenuation at GHz frequencies \cite{Erkip-14}. On the other hand, the use of mmWave has been advocated for fifth-generation (5G) systems in order to satisfy on ten-fold achievable data rate with respect to the previous mobile communication generation: in this scenario the use of extremely high-frequency bands becomes even more problematic due to the presence of multiple users and the need to design suitable multiple access schemes. From a paradigm based on orthogonality among users to maximize the sum rate -- mainly pursued in 4G systems -- the 5G trend is now shifting back to non-orthogonal multiple access (NOMA) schemes, better suited to heterogeneous devices envisioned in the 5G network, e.g., sensors and actuators of the Internet of Things (IoT) \cite{Gupta-15}. All these problems are further complicated by the use of a huge number of antennas, the massive \ac{MIMO} regime, again proposed for 5G systems in order to propel data rate and cell user density. In this context, a radical review of both beam-forming and scheduling is needed, taking into account also the need of keeping both the signaling exchange on air at the minimum (for a higher data-rate) and the signal processing complexity under control, especially on the terminal side. 

Focusing on a downlink scenario where a base station (BS) equipped with many antennas aims at serving multiple mobile terminals (MTs) on the mmWave band, many solutions are available in the literature for beam-forming and scheduling. For example, orthogonal user transmission can be considered in order to maximize the rate, thus (regularized) zero-forcing beamforming can be used at the BS, together with water-filling power allocation and scheduling \cite{Spencer-04, Yang-13}. Similar approaches include diagonalization and block-diagonalization of the channel \cite{Wong-03}. Indeed, it has been shown that in a massive MIMO regime, eigenbeamforming and matched filter precoders are optimal, and the transmit power can be made arbitrarily small \cite{Hoydis-13}. However, these results leverage orthogonality among users and have as target the maximization of the sum rate. These assumptions must be revised in 5G systems used also for device-to-device (D2D) communication in the IoT context, typically having \ac{FR} requirements: for these services the target becomes serving the largest number of users, providing them the required fixed data rate. In the NOMA context, two approaches have been proposed in the literature: superposition coding and successive interference cancellation. With the first approach users suffer from the interference but can still decode the intended signal as long as the signal to noise plus interference ratio (SNIR) is above a given threshold; with the latter approach the interfering signal is decoded and then canceled before the useful signal is decoded. For superposition coding a large literature is available, for the maximization of the weighted sum-rate and energy-efficient solutions (see \cite{Fang-16, Zte}  for surveys). In \cite{Ding-16} beamforming and power allocation have been designed using as metrics the sum rate and outage probability of MIMO-NOMA systems. In \cite{Shi-08} optimal beamforming and power allocation solutions have been derived for the weighted sum-rate maximization under a total power constraint and a max-min balancing of user rate under a total power constraint: in this work however no \ac{FR} user is present. A simple approach has been proposed in \cite{Nassar1, Nassar2}, where \ac{FR} and \ac{VR} users are allocated orthogonal resources, then using further orthogonal resources for the \ac{VR} users and non-orthogonal resources for the \ac{FR} users. In \cite{Naspatent} an scheduling considering also the pilot overhead needed for channel estimation, adapted to the user channels characteristics has been considered.

In this report we focus on the problem of scheduling and power allocation for the mixed user characteristics of \cite{Nassar1}: first, we formulate the problem of maximizing the weighted sum-rate of both \ac{FR} and \ac{VR} users, with constraints ensuring that active \ac{FR} users get the required rate. This turns out to be a mixed integer programming problem, where the integer variables dictate which \ac{FR} users are active, and continuous variables provide the power allocation for all users. The weights of the objective function provide a trade-off between the number of served \ac{FR} users and the resources allocated to variable users. We then exploit the peculiarity of mmWave channels which can be suitably represented as a sparse matrix in a dual domain \cite{Schniter}. The sparse channel is obtained by applying fixed discrete Fourier transform (DFT) beamformers at both the transmitter and the receiver, thus providing a low-complexity implementation. On the other hand, the sparsity of the resulting multiple access channel is exploited to simplify scheduling, first establishing an interference graph among users and then grouping users according to their orthogonality. In particular, the original scheduling problem is solved using a graph-coloring algorithm on the interference graph in order to select sub-sets of orthogonal \ac{VR} users. Two options are considered for \ac{FR} users: either they are chosen orthogonal to \ac{VR} users or they are allowed to interfere with the other set. A waterfilling algorithm is then used to allocate power to the \ac{FR} users, while the power needed to achieve the required rate is allocated to \ac{FR} users. The rest of the report is organized as follows. Firstly we derive the system model for the described scenario and give a mathematical description of the power allocation and scheduling problem. Secondly we propose three solutions: the direct power allocation and scheduling, a solution based on \ac{GC}, where both \ac{VR} and \ac{FR} users are clustered in different sets in order to eliminate interference, and a \ac{PGC} solution where only \ac{VR} users are clustered to eliminate interference. Lastly we analyse the performance of the proposed solutions in terms of achievable sum-rate when the optimization target moves from \ac{VR} users to \ac{FR} users and show how the number of antennas at the BS impacts on the proposed solutions.

\section{System Model}

We consider a massive \ac{MIMO} downlink transmission from a BS with a large number $N_{a}$ of antennas organized in a uniform planar array (UPA) to a set of $N_u$ single-antenna users, with $N_u << N_a$. Assuming an \ac{AWGN} flat-fading channel model among each antenna couple, the resulting $N_u \times N_a$ channel matrix is denoted as $\bm{H}$. For a transmission in the mmWave band, each BS-user link is described by $L$ paths, with $L$ typically being small. Moreover, considering the UPA configuration at the BS and the single antenna MT, as derived in \cite{Sayeed-02} the row channel vector $\bm{h}_k$ of user $k$ can be transformed into a virtual channel vector $\bm{h}_{{\rm v},k}$ by the transformation
\begin{equation}
\bm{h}_k = \bm{h}_{{\rm v},k} \bm{U}_t^*\,,
\end{equation}
where $\bm{U}_t$ is the $N_a \times N_a$ unitary DFT matrix. By stacking the rows $\bm{h}_{{\rm v}, k}$ into the matrix $\bm{H}_v$ we can write the  channel matrix as
\begin{equation}
 \bm{H}= \bm{H}_{\rm v}\bm{U}_t^*\,.
 \label{eq:hv}
\end{equation} 
When the number of antennas at the BS tends to infinity, the virtual channel matrix $\bm{H}_v$ becomes sparse, asymptotically having $L \cdot N_u$ non-zero entries associated to the $L$ paths of each one of the $N_u$ users. Also, when angles of departures are aligned with DFT direction \cite{Schniter} again $\bm{H}_v$ becomes sparse with $L \cdot  N_u$ non-zero complex entries, with probability density function
\begin{equation}
    p_{[\bm{H}_v]_{j,k}}(a) = (1-\eta)\delta(a)+\frac{\eta}{\pi \sigma^2_L}\exp\left(-\frac{|a|^2}{\sigma_L^2}\right)\,,
    \label{eq:chent}
\end{equation}
where $\eta=L/ N_a$ is the probability that a virtual channel matrix entry is different from zero, $\delta(\cdot)$ is the delta function and $\sigma_L^2$ is  the average channel gain.

Received signal is affected by complex \ac{AWGN}, independent at each user,  with zero mean and variance $\sigma^2$.

We consider two different types of users:
\begin{itemize}
\item $K$ \ac{VR} users having as target the maximization of their \ac{ASR};
\item $N$ \ac{FR} users aiming at being active, i.e., transmitting with a fixed data rate $R_{\rm fix}$, corresponding to fixed \ac{SINR} $\gamma_{\rm fix}$ with 
\begin{equation}
R_{\rm fix} = \log_2 (1+\gamma_{\rm fix})\,.
\end{equation}
\end{itemize}
The total number of users is $K+N = N_u$. A \ac{FR} user assumes only two possible states: transmission with a \ac{FR} imposed by the \ac{SINR} requirement or turn off. In particular, the first $K$ users are the \ac{VR} users, while next $N$ users are the \ac{FR} users.

We denote by $p_j$, $j=1, \ldots, N_u$ the power assigned to user $j$. We also collect into the $N_u$ column vector $\bm{p}$ all user powers. We consider a total power constraint, i.e.,
\begin{equation}
\label{powerconst}
 p_k \geq 0, \quad \sum_{k=1}^{N_u}p_k \leq 1\,.
\end{equation}

According to the beamformer adopted by the BS, as described in the following section, the resulting \ac{SINR} for user $k=1, \ldots, N_u$ is denoted as $\gamma_k(\bm{p})$, depending on the power allocation. For the beamforming and scheduling problem we aim at maximizing the weighted \ac{ASR}
\begin{equation}
    \mathcal{R}(\bm{p}) = \sum_{k=1}^{K} w_{k} \log_{2}(1+\gamma_{k}(\bm{p}))+ \sum_{k=K+1}^{K+N}\rho_k x_k R_{\rm fix}\,,
    \label{eq:sr}
\end{equation}
where $\rho$ is the weight of \ac{FR} users \ac{ASR}, $w_k$ is the weight of the \ac{ASR} of \ac{VR} user $k$ and $x_k$ is a binary variable which assumes value 1 if the considered \ac{FR} user is active and 0 otherwise.

\section{Beamforming and Scheduling}

Considering the properties of the virtual channel, and that it can be obtained by the inverse \ac{DFT} matrix applied at the BS, we consider as BS transmit beamformer matrix $\bm{U}_t$, therefore the equivalent channel seen by the users is the sparse virtual channel $\bm{H}_v$ having $N_a$ virtual transmit antennas and $N_u$ virtual receive antennas. We will use one transmit virtual antenna per user, and in particular for user $k$ we select in $\bm{h}_{{\bm v},k}$ the entry with largest gain, i.e.,
\begin{equation}
\ell_k = {\rm argmax}_i |[\bm{h}_{{\bm v},k}]_i|^2\,, 
\end{equation}
and we indicate with 
\begin{equation}
G_{k,k} = |[\bm{h}_{{\bm v},k}]_{\ell_k}|^2
\end{equation}
the corresponding gain. With this choice, user $k$ will suffer from the the interference of all other user signals, based on his virtual channel $\bm{h}_{{\bm v},k}$, i.e. the gain of the interference channel from user $j$ to user $k$ is 
\begin{equation}
G_{j,k} = |[\bm{h}_{{\bm v},k}]_{\ell_j}|^2\,,
\end{equation}
corresponding to the use of virtual antenna $\ell_k$ by user $j$ through the channel of user $k$. Therefore $N_u \times N_u$ matrix $\bm{G}$ having entries $G_{j,k}$ defines the useful gain and interference gain for all users. With this choice, user $k$ experiences a \ac{SINR} 
\begin{equation}
 \gamma_k(\bm{p}) = \frac{G_{k,k}p_k}{\sum_{j \neq k}G_{j,k}p_j+\sigma^2}\,.
 \label{eq:sinr}
\end{equation}

Let us consider the weighted ASR maximization problem with variables $p_k$ and $x_k$. Our aim is to select a subset of \ac{FR} users and to allocate power to both \ac{VR} users and active \ac{FR} users in order to maximize the weighted ASR. Let $\bm{x} = [x_{1}, \ldots, x_{K+N}]$. The problem can be modeled as
\begin{subequations}
\begin{equation}
\begin{split}
 \label{eq:max}
 \underset{\bm{p}, \bm{x}}{\text{max}}
  \sum_{k=1}^{K} w_k \log_{2}\bigg (1+\frac{G_{k,k} p_{k}}{\sum_{j=1, j \neq k}^{K+N}G_{j,k} p_{j} x_j +\sigma^2}\bigg )+ \\
   +\sum_{k=K+1}^{K+N}\rho_k R_{\rm fix}x_k 
 \end{split}
 \end{equation}
\text{s.t. (\ref{powerconst})} and
\begin{equation} 
\label{const1}
  \frac{G_{k,k} p_{k}}{\sum_{j=1, j \neq k}^{K+N}G_{j,k} p_{j} x_j +\sigma^2} \geq \gamma_{\rm fix} \quad k = K+1,...,K+N
\end{equation}
\begin{equation}
\quad x_k = 1 \quad k = 1,...,K
\end{equation}
\begin{equation}
\label{const2}
\quad x_k \in \{0,1\} \quad k = K+1,...,K+N
\end{equation}

\end{subequations}

This problem of joint power allocation and scheduling is modelled as a mixed-integer optimization problem, which belongs to the NP-hard class of problems. We will hence first analyse the  solution and then discuss two different methods that can reduce the computational complexity of the  solution while ensuring the same weighted ASR value.
 
\subsection{Direct solution}

In order to solve the mixed-integer programming problem we divide it into two problems: the problem of finding the set of active users (thus solving on $\bm{x}$) and the problem of allocating the power (thus solving on $\bm{p}$). In particular we exhaustively consider all possible sets of active \ac{FR} users, and for each set we solve the power allocation problem (if feasible) maximizing the weighted ASR. For each set of active \ac{FR} users $\bm{x}$ the resulting power allocation  problem is hence
\begin{equation}
\begin{split}
 \label{7a1}
 \underset{\bm{p} }{\text{max}}
  \sum_{k=1}^{K} w_k \log_{2}\bigg (1+\frac{G_{k,k} p_{k}}{\sum_{j=1, j \neq k}^{K+N}G_{j,k} p_{j} x_j +\sigma^2}\bigg )+ \\
   +\sum_{k=K+1}^{K+N}\rho_k R_{\rm fix}x_k 
 \end{split}
 \end{equation}
s.t. (\ref{powerconst}), (\ref{const1}) and (\ref{const2}).

We notice that the resulting optimization problem is non-convex due to the argument of the the log function of (\ref{7a1}) and we approximate  the target function to make it convex. In particular, we consider a high \ac{SINR} regime and perform the change of variables  $q_k = \log_2 p_k$, obtaining the convex maximization problem 
\begin{subequations}
\begin{equation}
\begin{aligned}
 &\underset{\bm{p}}{\text{max}}
 & & \sum_{k=1}^{K} w_k \log_{2}\bigg (\frac{G_{k,k} e^{q_{k}}}{\sum_{j=1, j\neq k}^{K+N}G_{j,k} e^{q_{j}} x_j +\sigma^2}\bigg )+
 \end{aligned}
 \end{equation}
 \begin{equation*}
   \sum_{k=K+1}^{K+N}\rho_k R_{\rm fix}x_k 
 \end{equation*}
\begin{equation} 
\text{s.t.}
\quad e^{q_k} \geq 0, \quad \sum_{k=1}^{K+N}e^{q_k} \leq 1
\end{equation}
\begin{equation} 
\label{const1}
  \frac{G_{k,k} e^{q_k}}{\sum_{j=1, j \neq k}^{K+N}G_{j,k} e^{q_j} x_j +\sigma^2} \geq \gamma_{\rm fix} \quad k = K+1,...,K+N.
\end{equation}
\end{subequations}
 The resulting maximization problem is hence convex and can be solved with standard optimization solutions. We then pick the set of active users for which the problem is feasible and provides that maximum weighted \ac{ASR}.

\subsection{\ac{GC} solution}

The second solution aims at reducing the computational complexity of the optimization problem by clustering the users in different groups, in which users transmit without interference. Then we transmit only to one set of non-interfering users, allocating powers in order to maximize the  weighted \ac{ASR}. The set of active users is the one that maximizes the weighted \ac{ASR} among all sets.

In order to perform the clustering operation we use the  interference matrix $\bm{G}$. We notice that, since the mmWave virtual channel results sparse, the resulting interference matrix computed over the matrix of the gains of the virtual channel $\bm{G}$ will have mostly small or zero off-diagonal entries. In particular, as the ratio $N_a/N_u$ of transmit and receive antennas tends to infinity,  the interference matrix itself will become sparse. 
Indeed, in sparse massive-\ac{MIMO} channels the channel vectors associated to different users become orthogonal as the number of antennas goes to infinity \cite{Gao}. Our system behaves in a similar manner. Considering the virtual channel model with $L$ non-zero paths and a growing number of antennas at the BS, we have that, asymptotically, users do not interfere. Therefore the interference matrix becomes a diagonal matrix and the graph colouring leads to all users labelled with the same colour.

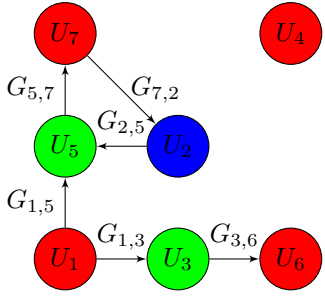
\begin{figure}
\centering
\begin{tikzpicture}
\centering
\tikzset{vertex/.style = {shape=circle,draw,minimum size=1.5em}}
\tikzset{edge/.style = {->,> = latex'}}
\node[vertex,  fill=red] (a) at  (1,1) {$U_1$};
\node[vertex, fill=blue] (b) at  (2.5,2.5) {$U_2$};
\node[vertex, fill=green] (c) at  (2.5,1) {$U_3$};
\node[vertex, fill=red] (d) at  (4,4) {$U_4$};
\node[vertex, fill=green] (a1) at (1,2.5) {$U_5$};
\node[vertex, fill=red] (a2) at (4,1) {$U_6$};
\node[vertex, fill=red] (a3) at (1,4) {$U_7$};
\draw[edge] (a) to node[above]{$G_{1,3}$} (c) ;
\draw[edge] (a) to node[left]{$G_{1,5}$} (a1);
\draw[edge] (a1) to node[left]{$G_{5,7}$} (a3);
\draw[edge] (c) to node[above]{$G_{3,6}$} (a2);
\draw[edge] (a3) to node[right]{$G_{7,2}$} (b);
\draw[edge] (b) to node[above]{$G_{2,5}$} (a1);
\end{tikzpicture}
\caption{Interference graph obtained from a particular realization of the interference matrix $\bm{G}$.}
\label{fig:graph}
\end{figure}

From the interference matrix we build the interference graph, i.e. a directed graph composed by a node for each user and with edges connecting interfering users: edges are directed from the interfering user toward the user suffering the interference. No edge between nodes $k$ and $n$ is present if and only if $G_{k,n} = 0$. Nodes are then clustered into groups of non-interfering users: this corresponds to color the graph, where each color represents a group and nodes with the same colors must not be connected by edges. We aim at finding the minimum number of colors needed for the graph, and we resort to the solution proposed in \cite{Segundo}. Fig. \ref{fig:graph} shows an example of interference graph obtained for a particular realization of the interference matrix, where, after graph colouring, each node (corresponding to user $U_k$)  has been assigned a colour red or green.
Since users having the same color do not interfere, assuming to transmit only to users of the same color the \ac{SINR} of user $k$ can be written as
\begin{equation}
\gamma_k = \frac{G_{k,k}p_k}{\sigma^2}\,,
\label{eq:snr}
\end{equation}
which coincides with the \ac{SNR}.

Let $\mathcal C$ the set of colors,  $\mathcal F_c$ be the set of indices of \ac{FR} users with color $c \in \mathcal C$, and let $\mathcal V_c$ the set of \ac{VR} users with color $c$. To select the subset of active \ac{FR} users (with the same color) we resort to the  exhaustive search approach, thus computing the weighted \ac{ASR} corresponding to each active set. For the \ac{FR} users the allocated power is 
\begin{equation}
p_k = \gamma_{\rm fix} \frac{\sigma^2}{G_{k,k}}x_k\,, \quad  k \in \mathcal F_c
\label{eq:poF}
\end{equation}
For \ac{VR} users since no interference is present, the optimal power allocation strategy is provided by the waterfilling algorithm 
\begin{subequations}
\begin{equation}
\begin{aligned}
 &\underset{ \{p_k, k \in \mathcal V_c\}}{\text{max}}
 & & \sum_{k\in \mathcal V_c} w_k \log_{2}\bigg (1+\frac{G_{k,k} p_{k}}{\sigma^2}\bigg) 
 \end{aligned}
 \end{equation}
\begin{equation}
\text{s.t.}
\quad p_k \geq 0, \quad \sum_{k\in \mathcal V_c}p_k \leq 1- \sum_{k \in \mathcal F_c} x_k p_k\,.
\end{equation}
\end{subequations}
The graph-coloring solution is summarized in  Algorithm 1.

\begin{algorithm}[]
\DontPrintSemicolon
 \KwData{${\bm{G}}$,$\sigma^2$, color set $\mathcal C$, indices sets $\mathcal V_c$ and $\mathcal F_c$, $c \in \mathcal C$.}
 \KwResult{$\bm{p}$, $\bm{x}$ }
 \For{$c \in \mathcal C$}{
   \For{all possible sets of \ac{FR} users $\bm{x}$}{
   $\forall k \in \mathcal F_k$ compute $p_k | (11b)$ is satisfied\;
    \For{$k \in \mathcal F_c$}{
 
  compute water-filling level $\lambda$ such that
 \begin{equation}
     \sum_{k \in \mathcal V_c}\bigg(\lambda-\frac{\sigma^2}{G_{k,k}}\bigg)^+ = 1- \sum_{k \in \mathcal F_c} x_k p_k
 \end{equation}
    $\mathcal R_c(\bm{x})$ = weighted ASR for $\bm{x}$\;    }
    
    }
    $\bm{x}_c = {\rm argmax}_{\bm{x}} \mathcal R_c(\bm{x})$
    }
    $c^* = {\rm argmax}_c \mathcal R_c$
    
    $\bm{p} = \bm{p}_{c^*}$
    
    $\bm{x} = \bm{x}_{c^*}$

 \caption{\ac{GC} Power allocation and Scheduling Algorithm}
\end{algorithm}

\subsection{\ac{PGC} solution}

The third solution applies grouping (i.e., \ac{GC}) only on \ac{VR} users, while the \ac{FR} users are allowed to interfere with the \ac{VR} users and among themselves. This allows a more flexible allocation of \ac{FR}  users, while at the same time due to their low-rate request we expect that the resulting interference on the \ac{VR} users is reduced. 

Therefore, in this case we apply the \ac{GC} only on the interference graph of the \ac{VR} users, while we must perform the exhaustive search of the the active \ac{FR} users among all \ac{FR} users. Therefore, for each color $c \in \mathcal C$ we must solve the optimal scheduling and power allocation problem for the set of \ac{VR} users $\mathcal V_c$ and for all \ac{FR} users. With respect to the global direct approach we have reduced the set of \ac{VR} users, but on the other hand we must run the optimization algorithm for each color.

\section{Numerical Results}

We  present now  numerical results obtained with the different solutions described in previous sections. We consider $N_u=7$ users where $K=4$ are \ac{VR} users and $N=3$ are \ac{FR} users. We assume  that the virtual channel matrix $\bm{H}_v$ is composed by $L=100$ non-zero entries and that they are distributed as in (\ref{eq:chent}) and   the SINR requirement for \ac{FR} users is $5$ dB while \ac{AWGN} leads to an average  \ac{SNR} of 17 dB.

\begin{figure}[h]
 \centering
 \includegraphics[width=1\hsize]{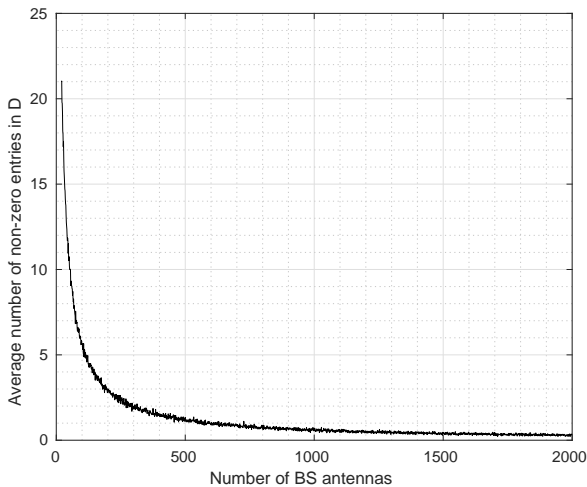}
        \caption{Average number of non-zero entries in $\bm{D}$ vs. number of antennas at the BS.}
        \label{fig:as}
\end{figure}

We start verifying the assumption that the interference matrix is sparse, due to the large number of BS antennas. Let us consider  matrix $\bm{D}$, defined as the matrix containing off-diagonal elements of $\bm{G}$, i.e., $\bm{D} = \bm{G} - {\rm diag}(\bm{G})$. Fig. \ref{fig:as} shows the average number of non-zero entries of $\bm{D}$ vs the number of antennas at the BS. We observe that the average number of non-zero entries sharply decreases as the number of antennas increases, and at $N_a =200$ we have on average about 3 non-zero off-diagonal entries. In the following we consider $N_a =200$.

\begin{figure}
 \centering
 \includegraphics[width=1\hsize]{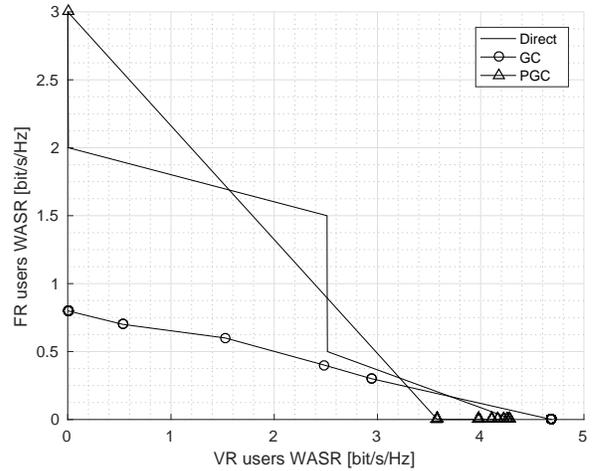}
        \caption{\ac{FR} vs. \ac{VR} users \ac{ASR} for different $\rho$ values.}
        \label{fig:all}
\end{figure}

We now consider the scheduling and power allocation algorithms described in the paper. The weight $w_k$ has been chosen to be equal to $1-\rho$ for each \ac{VR} user, i.e. $w_k = 1-\rho \quad \forall k=\{1,...,K\}$ and $\rho \in \{0,1\}$. With this choice we do not distinguish among the users in  the two sets, but only use $\rho$ to balance the resources uses for \ac{VR} and \ac{FR} users. In particular, when $\rho \approx 0$, we maximize the \ac{ASR} of \ac{VR} users, whereas for $\rho \approx 1$ we maximize the \ac{ASR} of \ac{FR} users. Therefore we consider as performance metric the average \ac{ASR} for both \ac{VR} and \ac{FR} users. Fig. \ref{fig:all} shows the average (over various channel realizations) ASR vs. $\rho$, for the different scheduling and power allocation solutions.  The direct algorithm is denoted as {\tt direct}, and then we have the \ac{GC} and \ac{PGC} algorithms. As we expect, with increasing values of $\rho$, VR users ASR decreases while FR users ASR increases. Note also that even if $\rho = 1$ we still have a non-zero ASR for \ac{VR} users, since all \ac{FR} are served and the remaining power is allocated to \ac{VR} users.

We notice that the direct solution to power allocation and scheduling is well approximated by both \ac{GC} and \ac{PGC} solutions for VR users, whereas, for FR users, the \ac{PGC} solution is the best performing. Worse ASR performance for FR users in GC solution are due to the fact that VR users could be labelled with different colours and hence can not transmit together. Notice that the GC solution for VR users outperforms the direct solution. This is due to the fact that even if the maximization problem is focused on FR users ASR, they could be labelled with different colours and hence in GC solution they can not transmit all together, resulting in a power allocation for VR users with higher values than direct solution. This results in higher ASR values for VR users when $\rho$ tends to 1.

\section{Conclusions}
In this work we considered a mmWave massive \ac{MIMO} system and we solved the problem of power allocation and users scheduling when two different sets of users transmit: \ac{VR} users and \ac{FR} users with a required \ac{SINR} level for transmission. We analysed the direct solution and proposed two algorithms, GC and PGC, that present similar ASR performance for both \ac{VR} users and \ac{FR} users while reducing the computational complexity. Then we analysed the performance of all the proposed solutions in terms of ASR when the weight $\rho$ (and hence $w_k$) assigned to \ac{VR} users changes and showed that the proposed algorithms present the same performance obtained with the direct solution. We also showed that with an increasing number of antennas at the BS the interference matrix becomes diagonal and hence that the low-complexity GC solution achieves optimal results.

\balance


\end{document}